\documentclass[review]{elsarticle}

\usepackage{lineno}
\modulolinenumbers[5]

\journal{Applied Mathematical Modelling}

\usepackage{amsmath,amssymb}
\usepackage{array,tabularx,tabulary,booktabs} %
	\usepackage{longtable}  
 	\usepackage{multirow} 
	\usepackage{makecell}
\newcommand{\specialcell}[2][c]{%
	\begin{tabular}[#1]{@{}l@{}}#2\end{tabular}}

\usepackage{siunitx} 
\usepackage[hidelinks]{hyperref} 
\usepackage{cleveref} 

\begin{document}

\begin{frontmatter}

\title{Semi-implicit finite-difference method \\with predictor-corrector algorithm \\for solution of diffusion equation with nonlinear terms}


\author[mymainaddress,mysecondaryaddress,mythirdaddress]{V.P. Lipp}
\ead{vladimir.lipp@desy.de}

\author[mymainaddress]{B. Rethfeld}

\author[mysecondaryaddress]{M.E. Garcia}

\author[mymainaddress,mysecondaryaddress]{D.S. Ivanov}

\address[mymainaddress]{Department of Physics and OPTIMAS Research Center,\\Technical University of Kaiserslautern, 67663 Kaiserslautern, Germany}
\address[mysecondaryaddress]{Theoretical Physics (FB10), University of Kassel, 34132 Kassel, Germany}
\address[mythirdaddress]{Center for Free-Electron Laser Science, Deutsches Elektronen-Synchrotron DESY, Notkestr. 85, 22607 Hamburg, Germany}

\begin{abstract}
We present a finite-difference integration algorithm for solution of a system of differential equations containing a diffusion equation with nonlinear terms. The approach is based on Crank-Nicolson method with predictor-corrector algorithm and provides high stability and precision. Using a specific example of short-pulse laser interaction with semiconductors, we give a detailed description of the method and apply it for the solution of the corresponding system of differential equations, one of which is a nonlinear diffusion equation. The calculated dynamics of the energy density and the number density of photoexcited free carriers upon the absorption of laser energy are presented for the irradiated thin silicon film. The energy conservation within \SI{0.2}{\%} has been achieved for the time step $10^4$ times larger than that in case of the explicit scheme, for the chosen numerical setup. We also present a few examples of successful application of the method demonstrating its benefits for the theoretical studies of laser-matter interaction problems.\end{abstract}


\end{frontmatter}

\section{Introduction}
Many phenomena occurring in nature for their investigation can be described via mathematical models based on time-dependent nonlinear diffusion equations \cite{wu2001nonlinear}. Examples include genetics \cite{mckane2007singular,bower2004computational}, image processing \cite{aubert2006mathematical}, quantum mechanics \cite{nagasawa2012schrodinger}, and laser-material interactions \cite{bauerle2011laser}. Although during the last decades big effort has been undertaken to find efficient numerical schemes for solution of the corresponding mathematical problem, some of the applications are still a challenging task. Specifically, the efforts in the model implementation as well as their demands on the computational power during processing can substantially hinder the theoretical interpretation of the investigated problem. In this work, we consider an application of the nonlinear parabolic diffusion equation to describe the response of solids to an ultrashort laser pulse irradiation. Apart from the insights into the material structure, this topic is important for the description of laser machining \cite{liu1997laser,chichkov1996femtosecond,gattass2008femtosecond} and nanostructuring experiments \cite{chou2002ultrafast,le2011generation,huang2008nucleation} with applications in Bio- \cite{stratakis2011biomimetic} and IT-technologies \cite{mathis2012micromachining,bhuyan2010high}. For metals, the problem may be mathematically formulated in the form of frequently used Two-Temperature Model (TTM) \cite{anisimov1974electron}, whereas for semiconductors a similar TTM-like approach has been proposed \cite{van1987kinetics}. The latter is based on the system of partial differential equations, reflecting the conservation laws in the atomic subsystem of a solid and its electronic subsystem. Though it is relatively simple to apply an explicit finite-difference numerical scheme to solve such systems in metals \cite{ivanov2003combined} or semiconductors \cite{chen2005numerical}, the corresponding stability criteria demand the integration time steps to be small, causing high computational costs as a result. The main restriction on the time step often comes from the nonlinear diffusion equation describing the carrier heat conduction process \cite{isaacson2012analysis}. One of the possibilities to increase the time step of diffusion equation is to use implicit or semi-implicit integration schemes. For instance, the Crank-Nicolson semi-implicit scheme \cite{crank1947practical,cebeci2002convective} provides unconditionally stable solution when applied to linear diffusion equations. However, this approach is not directly applicable when nonlinear terms play an important role. 

In this work, we present a semi-implicit finite-difference method for the solution of a system of differential equations, one of which is a diffusion equation with nonlinear terms, and apply it to model short laser pulse interaction with semiconductors on the example of silicon. The presented approach is based on Crank-Nicolson method with predictor-corrector algorithm and provides high stability and precision. It has been already successfully applied for the investigation of ultrashort laser interaction with metals \cite{ivanov2014molecular} and semiconductors \cite{ramer2014laser,lipp2014atomistic}. 
\Cref{subsec:modeltosolve} is devoted to the continuum TTM-like model for semiconductors. We describe the theoretical model and present the system of equations where a nonlinear diffusion equation results in strong restriction on the time step for the explicit integration algorithm. In \cref{sec:implicit_scheme}, we give a detailed description of semi-implicit numerical solution scheme, which was modified with predictor-corrector algorithm to account for the nonlinearity in the diffusion equation. Further, in \cref{sec:sol-alg-results}, the calculation results for a particular set of parameters are presented and the energy conservation versus the applied iteration time step is investigated. \Cref{sec:sol-alg-discuss} mentions the existing works, in which this approach has been successfully utilized, and suggests possible improvements for the application of the presented approach in two-dimensional (2D) and three-dimensional (3D) case. Finally, in \cref{sec:conclude} we give a summary of our results.

\section{Model description}
\label{subsec:modeltosolve}

Here we present the full set of nonlinear differential equations for the continuum description of electron density and electron/phonon energy density dynamics in silicon under ultrashort laser irradiation. For the derivation of the following expressions we refer to \cite{van1987kinetics}. Due to laser pulse irradiation (in this example Ti:Sapphire laser at \SI{800}{\nano\meter} wavelength), free carriers are generated in the material, electrons in the conduction band and holes in the valence band, by one- and two-photon absorption processes. Both types of carriers are assumed to quickly equilibrate in the corresponding parabolic bands. We assume the Dember field prevents charge separation, consequently the two types of carriers move together \cite{young1982ambipolar}. To each of them, we apply separate Fermi-Dirac distributions with different chemical potentials, $\phi_e$ and $\phi_h$ for the electrons and holes respectively, but with shared carrier density $n$ and temperature $T_e$ (two-chemical-potentials model):
\begin{equation}\label{eq:FD-m}
f_c\left( E\right) =\frac{1}{\exp\left( \frac{\pm\left( E-\phi_c\right)}{k_B T_e} \right) +1},
\end{equation}
where subscript $c$ stands as $e$ for electrons and $h$ for holes; the $+$ sign is associated with electrons and the $-$ sign with holes. The reduced chemical potentials are defined as follows:
\begin{equation}\label{eq:eta_c-m}
\eta_e=\frac{\phi_e-E_C}{k_B T_e}\text{ and } \eta_h=\frac{E_V-\phi_h}{k_B T_e},
\end{equation}
where $E_C$ and $E_V$ are the conduction and valence band energy levels respectively, so the energy gap is $E_g = E_C-E_V$. The integration of the carrier distribution functions over the energy leads to the expressions for the carrier density (parabolic bands are assumed):
\begin{equation}\label{eq:n_c-m-m}
n=2\left( \frac{m_c^* k_B T_e}{2\pi \hbar^2}\right)^{\frac{3}{2}} F_{\frac{1}{2}}\left( \eta_c\right).
\end{equation}
The Fermi-Dirac integral is defined as:
\begin{equation}\label{eq:FD_integral-m}
F_\xi\left( \eta_c\right) =\frac{1}{\Gamma\left(\xi+1 \right) } \displaystyle\int_{0}^{\infty} \! \frac{x^\xi}{1+\exp\left( x-\eta_c\right) } \, \mathrm{d}x.
\end{equation}
The carrier current is the sum of contributions from the electrons and the holes:
\begin{equation}\label{eq:Ju-m}
\begin{split}
\vec{J}=-D& \Bigg[ \nabla n + \frac{n}{k_B T_e} \left[ H_{-\frac{1}{2}}^\frac{1}{2}\left( \eta_e \right) +  H_{-\frac{1}{2}}^\frac{1}{2}\left( \eta_h \right)  \right]^{-1} \nabla E_g  +\\
& \frac{n}{T_e} \left[ 2\frac{H_{0}^1\left( \eta_e \right) +H_{0}^1\left( \eta_h \right)}{H_{-\frac{1}{2}}^\frac{1}{2}\left( \eta_e \right) +  H_{-\frac{1}{2}}^\frac{1}{2}\left( \eta_h \right) } -\frac{3}{2}\right] \nabla T_e \Bigg],
\end{split}
\end{equation}
where $H_\zeta^\xi\left( \eta_c \right) \equiv F_\xi \left( \eta_c \right) / F_\zeta \left( \eta_c \right) $ and the ambipolar diffusion coefficient is:
\begin{equation}\label{eq:D-m}
D=\frac{k_B T_e}{q_e}\frac{\mu_e \mu_h H_\frac{1}{2}^0 \left( \eta_e \right) H_\frac{1}{2}^0 \left( \eta_h \right)}{\mu_e H_\frac{1}{2}^0\left( \eta_e \right)+\mu_h H_\frac{1}{2}^0\left( \eta_h \right)}
 \left[ H_{-\frac{1}{2}}^\frac{1}{2}\left( \eta_e \right) +  H_{-\frac{1}{2}}^\frac{1}{2}\left( \eta_h \right)\right]
\end{equation}
with $q_e$ the elementary charge.
Ambipolar energy flow is the sum of diffusion and thermal energy currents inside the carrier subsystem and can be written as:
\begin{equation}\label{eq:W-m}
\vec{W}=\left\lbrace E_g+2k_B T_e \left[H_0^1\left(\eta_e \right) + H_0^1\left( \eta_h\right)  \right]  \right\rbrace \vec{J} - \left(\kappa_e + \kappa_h \right) \nabla T_e.
\end{equation}

The dynamics of semiconductors under the irradiation of ultrashort laser pulses can be modeled with the system of three continuum equations \cite{van1987kinetics,lietoila1982computer}: continuity equation for free carrier density and two coupled energy balance equations, one for carriers and one for atoms:
\begin{equation}\label{eq:dn_dt-m}
\frac{\partial n}{\partial t}+\nabla \cdot \vec{J} = S_n - \gamma n^3 + \delta\left( T_e \right)n ,
\end{equation}
\begin{equation}\label{eq:du_dt-m}
\frac{\partial u}{\partial t}+\nabla \cdot \vec{W} = S_u - \frac{C_{e-h}}{\tau_{ep}}\left( T_e-T_a \right),
\end{equation}
\begin{equation}\label{eq:dTa_dt-m}
C_a\frac{\partial T_a}{\partial t}=\nabla \cdot \left( k_a\nabla T_a\right) + \frac{C_{e-h}}{\tau_{ep}}\left( T_e-T_a \right).
\end{equation}
The meanings of symbols in \cref{eq:dn_dt-m,eq:du_dt-m,eq:dTa_dt-m} are the following: $S_n$ is the source of new carriers (excitation rate of new carriers by the laser), $u$ is the carrier energy density, $S_u$ describes the energy source (rate of laser energy absorption), $C_{e-h}$ is specific heat capacity of the electron-hole pairs, $T_a$ is atomic temperature. The terms on the right hand side of \cref{eq:dn_dt-m} are responsible for the laser energy absorption, Auger recombination, and impact ionization, consequently. \Cref{eq:du_dt-m} describes the energy balance in the photoexcited electron-hole pairs and is a nonlinear diffusion equation. The terms on the right hand side are responsible for the laser energy absorption and the coupling to the lattice. The last equation (\ref{eq:dTa_dt-m}) describes the energy balance in the atomic subsystem. The parameters used in the calculations as well as the meanings of other symbols are presented in \cref{Tab:ModelParam-m}.

\begin{table}
	\caption{Model parameters.} \label{Tab:ModelParam-m}
	\begin{tabularx}{\textwidth}{|l|X|r|}
		\hline 
		Parameter name & Value &  Citation \\ 
		\hline 
		\specialcell{Initial carrier density} & \SI[mode=text]{1\times 10^{16}}[$n_0=$ ]{m^{-3}} & \cite{sproul1991improved}  \\
		\hline
		\specialcell{Initial lattice and \\ carrier temperature} & \SI[mode=text]{300}[$T_0=$ ]{\kelvin} & \\
		\hline
		Lattice specific heat & $C_a = 1.978\times10^6 + 3.54\times10^2T_a - 3.68\times10^6/T_a^2$ , J/(m$^3$K) ($T_a$ in K) & \cite{wood1981macroscopic} \\
		\hline
		\specialcell{Lattice thermal \\ conductivity} & \specialcell{$k_a = 1.585 \times 10^5 \times T_a^{-1.23}$, W/(m$\cdot$K) \\($T_a$ in K)} & \cite{wood1981macroscopic} \\
		
        \hline
		\specialcell{Carrier thermal \\ conductivity} & \specialcell{$k_e = k_h = -3.47\times10^{18} +$ \\ $4.45\times10^{16} T_e$, eV/(s m K) } & \cite{agassi1984phenomenological} \\
		\hline
	    Indirect band gap & 
	   \specialcell{$E_g = 1.170 - 4.73\times10^{-4}T_a^2/(T_a+ $\\$ 636) - 1.5\times10^{-10}n^{1/3}$ \\
	    if $1.170 - 4.73\times10^{-4} T_a^2/(T_a+$ \\ $636) - 1.5\times10^{-10}n^{1/3} \geq 0$ \\ and $0$ otherwise, eV \\ ($T_a$ in K, $n$ in m$^{-3}$)}
	     &
	    \specialcell{\cite{thurmond1975standard} \\ \cite{vankemmel1993unified}} \\	
	    \hline
	    \specialcell{Interband absorption \\
	    (taken from \SI{694}{\nano\meter} laser)} & $\alpha = 1.34\times10^5\exp \left( T_a/427\right)$, m$^{-1}$ & 
	    \cite{jellison1982optical} \\
	    
	    \hline
	    Two-photon absorption & \SI[mode=text]{15}[$\beta=$ ]{cm/GW} & \cite{lipp2014atomistic}
	    \\
	    \hline
	    Reflectivity & \specialcell{$R = 0.329 + 5\times10^{-5}(T_a - 300)$ \\($T_a$ in K)} & \cite{jellison1983optical} \\
	    
	    \hline
	    \specialcell{Auger recombination \\ coefficient} & $\gamma = 3.8\times10^{-43}$, m$^6$/s & \cite{dziewior1977auger}\\
	    
	    \hline
	    \specialcell{Impact ionization \\ coefficient} & \specialcell{$\delta = 3.6\times10^{10}\exp\left( -1.5E_g/k_BTe\right) $, \\s$^{-1}$}  & \cite{geist1983transition} \\
	    
	    \hline
	    \specialcell{Free-carrier absorption \\ cross section} & \specialcell{$\Theta = 2.91\times10^{-22} T_a/300$, m$^2$ \\(Ta in K) } & \cite{meyer1980optical} \\
	    
	    \hline
	    \specialcell{Electron-phonon \\ relaxation time} & \specialcell{$\tau_{e-p} = 0.5\times10^{-12} \left[ 1 + \right. $ \\$ \left. n/(2\times10^{27}) \right] $, s ($n$ in m$^{-3}$)} & \cite{agassi1984phenomenological} \\
	    \hline
\end{tabularx} 
\end{table}

\begin{table*}
	\begin{tabularx}{\textwidth}{|l|X|r|}
	    
	    \hline
	    Electron effective mass & $m_e^* = 0.36m_e$ & \cite{ioffe} \\
	    
	    \hline
	    Hole effective mass & $m_h^* = 0.81m_e$ & \cite{ioffe} \\
	    
	    \hline
	    \specialcell{Mobility of electrons \\ (taken at \SI{1000}{K})} & $\mu_e=$ \SI[mode=text]{0.0085}{m^2/V$\cdot$ s} & \cite{meyer1980optical} \\
	    
	    \hline
\specialcell{Mobility of holes \\ (taken at \SI{1000}{K})} & $\mu_h=$ \SI[mode=text]{0.0019}{m^2/V$\cdot$ s} & \cite{meyer1980optical}	    \\
		\hline
\end{tabularx} 
\end{table*}

The total energy of electron-hole pairs consists of the energy gap and the kinetic energy of electrons and holes (taking into account the Fermi statistics),
\begin{equation}\label{eq:u-m}
u=n E_g \left( n,T_e\right) + \frac{3}{2} n k_B T_e \left[ H_{\frac{1}{2}}^{\frac{3}{2}}\left( \eta_e\right)  + H_{\frac{1}{2}}^{\frac{3}{2}}\left( \eta_h\right)\right] .
\end{equation}

Let us rewrite \cref{eq:dn_dt-m,eq:du_dt-m,eq:dTa_dt-m} into more convenient form. Though it is written in the conservative form (which prevents the accumulation of numerical errors, providing the exact energy conservation in case of numerical solution) it is not convenient to solve, since equation (9) includes both variables $T_e$ and $u$. One can rewrite it with respect to $n_e$, $T_a$, and $T_e$ for a more handy numerical form, as it was suggested in \cite{van1987kinetics}. To do so, we have to note that the carrier specific heat capacity is $C_{e-h} = \partial u/\partial T_e |_n$; using \cref{eq:u-m} we can therefore obtain:
\begin{equation}
\label{eq:C_eh-m}
\begin{split}
C_{e-h}=&\frac{3}{2} nk_B \Bigg[  H_{\frac{1}{2}}^{\frac{3}{2}} \left(\eta_e \right) + H_{\frac{1}{2}}^{\frac{3}{2}} \left(\eta_h \right) + \\
& T_e \frac{\partial \eta_e}{\partial T_e} \left[ 1 - H_{\frac{1}{2}}^{\frac{3}{2}}\left( \eta_e \right)
H_{\frac{1}{2}}^{-\frac{1}{2}}\left( \eta_e \right)  \right] + 
T_e \frac{\partial \eta_h}{\partial T_e} \left[ 1 - H_{\frac{1}{2}}^{\frac{3}{2}}\left( \eta_h \right)
H_{\frac{1}{2}}^{-\frac{1}{2}}\left( \eta_h \right)  \right] \Bigg].
\end{split}
\end{equation}
Further, calculating $\frac{\partial u}{\partial t}$ from \cref{eq:u-m} and substituting it into \cref{eq:du_dt-m}, we arrive at the diffusion-like equation for the temperature of electron-hole pairs:
\begin{equation}\label{eq:dTe_dt-m}
\begin{split}
C_{e-h}& \frac{\partial T_e}{\partial t} = S_u - \nabla \cdot \vec{W} - \frac{C_{e-h}}{\tau_{e-p}} \left( T_e-T_a \right) - \\
&\frac{\partial n}{\partial t} \left\lbrace E_g+\frac{3}{2}k_B T_e
\left[ H_{\frac{1}{2}}^{\frac{3}{2}} \left( \eta_e \right) +
H_{\frac{1}{2}}^{\frac{3}{2}} \left( \eta_h \right) \right] -
n \left( \frac{\partial E_g}{\partial n} \frac{\partial n}{\partial t} + \frac{\partial E_g}{\partial T_a} \frac{\partial T_a}{\partial t} \right)  \right\rbrace - \\
&\frac{3}{2} k_B T_e n \frac{\partial n}{\partial t}
\left\lbrace \left[ 1- H_{\frac{1}{2}}^{\frac{3}{2}} \left( \eta_e \right) H_{\frac{1}{2}}^{\frac{3}{2}} \left( \eta_e \right) \right]
\frac{\partial \eta_e}{\partial n} +
\left[ 1- H_{\frac{1}{2}}^{\frac{3}{2}} \left( \eta_h \right) H_{\frac{1}{2}}^{-\frac{1}{2}} \left( \eta_h \right) \right] 
\frac{\partial \eta_h}{\partial n}  \right\rbrace 
\end{split}
\end{equation}

In order to present an example of the model application, we use the following source terms. The rate of free carriers density growth $S_n$ and the corresponding rate of their energy increase $S_u$ are given by:
\begin{equation}\label{eq:S_n-m}
S_n=\frac{\alpha I_{abs}\left( \vec{r},t\right) }{\hbar \omega} + \frac{\beta I_{abs}^2\left( \vec{r},t\right) }{2\hbar \omega},
\end{equation}
\begin{equation}\label{eq:S_u-m}
S_u=\alpha I_{abs}\left( \vec{r},t\right) + \beta I_{abs}^2\left( \vec{r},t\right) + \Theta n I_{abs}\left( \vec{r},t\right).
\end{equation}
In the last two equations, first and second terms on the right hand side represent the influence of one- and two-photon absorption, respectively, and the third term in the second equation represents the laser energy absorption by the excited free carriers.

Because of typically large laser spot size, as compared with the lateral sizes of the computational setup, the radial intensity distribution can be neglected until the radial diffusion starts to strongly contribute to the results. Therefore it can be sufficient to describe absorption and transport only in the direction of laser beam propagation at the center of the laser spot. Consequently, one-dimensional (1D) heating problem is analyzed in this work. The laser is focused on a material surface. The corresponding form of laser intensity at the surface ($z = 0$) in this case is:
\begin{equation}\label{eq:I_abs-m}
I_{abs}\left( 0,t\right) = \left( 1-R\left(T_a\right) \right) \sqrt{\frac{\varsigma}{\pi}}\frac{\Phi_{inc}}{t_p} \exp \left( -\varsigma \left[ \left( t-3t_p \right)/t_p  \right]^2  \right),
\end{equation}
where $\Phi_{inc}$ is the incident fluence, $\varsigma = 4\ln2$, and $R(T_a)$ is the reflectivity function (see \cref{Tab:ModelParam-m}). In this work, to prescribe the demanded incident fluence, the center of Gaussian pulse is shifted in time from the initial time $t = 0$ to \num{3} pulse duration times, $3\,t_p$, which in turn is defined as the pulse width at the half of maximum.

The dependence of the laser pulse intensity, $I_{abs}$, on depth can be found upon the solution of differential equation of the attenuation process:
\begin{equation}\label{eq:dI_dz-m}
\frac{dI_{abs}}{dz}=-\alpha I_{abs}\left( z,t\right) -\beta I_{abs}^2\left( z,t\right) -\Theta n I_{abs}\left( z,t\right) ,
\end{equation}
where $z$ is the depth into sample; the terms on the right side are responsible for one-, two-photon absorption, and for the free-carrier absorption processes consequently.

Thus, from the system of equations \cref{eq:dn_dt-m,eq:dTe_dt-m,eq:dTa_dt-m}, we can fully determine the dynamics of 
$n$, $T_e$, and $T_a$ in 1D using the following initial and boundary conditions, suitable for a free standing film:
\begin{equation}\label{eq:boundary-m}
\begin{split}
T_a\left(z,0 \right) &= T_e\left( z,0 \right) = \SI{300}{K}, \\
n\left( z,0\right) &= n_{eq} = \SI{1\times 10^{16}}{m^{-3}}   , \text{ ref.  \cite{sproul1991improved}} ,\\
J\left( 0,t\right) &= J\left( L,t\right) = 0, \\
W\left( 0,t\right) &= W\left( L,t\right) = 0, \\
k_a \frac{\partial T_a}{\partial z} \left( 0,t\right) &= k_a \frac{\partial T_a}{\partial z} \left( L,t\right) = 0,
\end{split}
\end{equation}
where $L$ is the thickness of the sample.

Owing to its similarity with an ordinary well-known TTM model \cite{anisimov1974electron}, but with an additional equation for free carriers density $n$, here and later we will refer to the described approach as \textit{n}TTM model, as it was suggested in \cite{ramer2014laser}.

\section{Numerical solution scheme}
\label{sec:implicit_scheme}

\begin{figure}
	\centering
	\includegraphics[width=\textwidth]{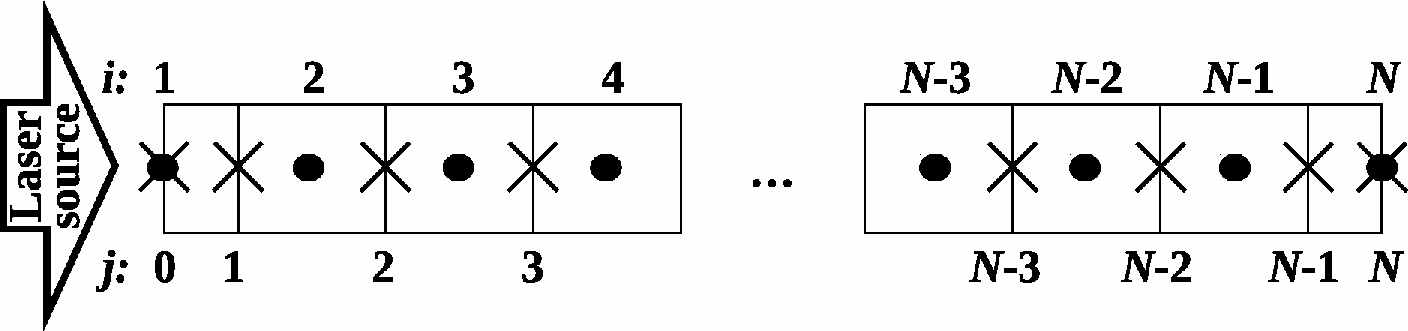}
	\caption{The finite-difference grid mesh for the solution of the \textit{n}TTM model equations. Symbol "$\bullet$" indicates the grid points for $n$, $Te$, $T_a$, and $E_g$ ($i=1, 2, ..., N$); symbol "$\times$" indicates the grid points for $\vec{J}$, $\vec{W}$, $D$, and  $k_a\frac{\partial T_a}{\partial z}$ ($j = 1, 2, ..., N+1$).} 
	\label{fig:nTTMgrid}
\end{figure}

In order to solve the described system of \cref{eq:dn_dt-m,eq:dTe_dt-m,eq:dTa_dt-m}, we use the finite difference grid mesh
sketched in \cref{fig:nTTMgrid}. Sample is divided into cells according to the scheme, and the local thermodynamic parameters are calculated in each cell. The spatial derivatives of 
$n, T_e, T_a, J, W, k_a\frac{\partial T_a}{\partial z}$, and $E_g$
at the interior points are approximated with the central differences, and those at the boundaries are evaluated with the first-order approximation.
\Cref{eq:dn_dt-m,eq:dTa_dt-m} are solved explicitly
($T\equiv T_e$):
\begin{equation}\label{eq:19-m}
\frac{n_i^{k+1}-n_i^{k}}{\Delta t}+\frac{J_i^k-J_{i-1}^k}{\Delta x}=(S_n)_i^k-\gamma(n_i^k)^3+\delta_i^kn_i^k,
\end{equation}
\begin{equation}\label{eq:20-m}
\begin{split}
(C_a)_i^k\frac{(T_a)_i^{k+1}-(T_a)_i^k}{\Delta t}=&\frac{1}{(\Delta z)^2}\left[(k_a)^k_{i+\frac{1}{2}}(T_{i+1}^k-T_{i}^k)-(k_a)^k_{i-\frac{1}{2}}(T_{i}^k-T_{i-1}^k)\right] \\&+\frac{(C_{e-h})_i^k}{(\tau_{e-p})_i^k}\left[T_i^k-(T_a)_i^k\right] ,
\end{split}
\end{equation}
where index $i$ is connected to the cell number (see \cref{fig:nTTMgrid}) and $k$ to the moment of time.

Therefore before solving \cref{eq:dTe_dt-m} we already have the information about $n^{k+1}$ and $(T_a)^{k+1}$. The
approach is based on the Crank-Nicolson semi-implicit scheme \cite{crank1947practical,cebeci2002convective}. \Cref{eq:dTe_dt-m} can be rewritten in the following finite-difference form:
\begin{equation}\label{eq:21-m}
\frac{T_i^{k+1}-T_i^k}{\Delta t}=(1-\psi)f_i^k+\psi f_i^{k+1} .
\end{equation}
The right-hand side contains parameter $\psi$, which can be \num{0} for explicit scheme, \num{1} for implicit, and $\frac{1}{2}$ for semi-implicit. The function $f_i^k$ can be found from:
\begin{equation}\label{eq:22-m}
\begin{split}
(C_{e-h})_i^k f_i^k=&(S_u)_i^k-\frac{W_i^k-W_{i-1}^k}{\Delta z}- 
\frac{(C_{e-h})_i^k}{(\tau_{e-p})_i^k}\left[T_i^k-(T_a)_i^k\right]- \\
&\left(\frac{\partial n}{\partial t}\right)_i^k\left\lbrace (E_g)_i^k+\frac{3}{2}k_BT_i^k 
\left[H_\frac{1}{2}^\frac{3}{2}(\eta_e)+H_\frac{1}{2}^\frac{3}{2}(\eta_h)\right]_i^k \right\rbrace- \\
&n_i^k\left\lbrace \left(\frac{\partial E_g}{\partial n}\right)_i^k \left(\frac{\partial n}{\partial t} \right)_i^k+\left(\frac{\partial E_g}{\partial T_a}\right)_i^k\left(\frac{\partial T_a}{\partial t}\right)_i^k   \right\rbrace- \\
&\frac{3}{2}k_BT_i^kn_i^k\left(\frac{\partial n}{\partial t} \right)_i^k \times
\left\lbrace \left(\frac{\partial\eta_e}{\partial n}
\right)_i^k \left[1-\left(H_{\frac{1}{2}}^{\frac{3}{2}}
(\eta_e)\right)_i^k \left(H_{\frac{1}{2}}^{-\frac{1}{2}}
(\eta_e)\right)_i^k\right]+ \right. \\
&\left. \left(\frac{\partial\eta_h}{\partial n} \right)_i^k
\left[1-\left(H_\frac{1}{2}^\frac{3}{2}(\eta_h)\right)_i^k
\left(H_\frac{1}{2}^{-\frac{1}{2}}(\eta_h)\right)_i^k
\right] 
\right\rbrace ,
\end{split}
\end{equation}
where $W_i^k$ is defined according to \cref{fig:nTTMgrid} (between the cells) and \cref{eq:W-m}:
\begin{equation}\label{eq:23-m}
\begin{split}
W_i^k=\left\lbrace (E_g)_{i+\frac{1}{2}}^k+ \right.
2k_BT_{i+\frac{1}{2}}^k[H_0^1(\eta_e)+H_0^1(\eta_h)]_{i+\frac{1}{2}}^k\times J_i^k-\\(k_e+k_h)_{i+\frac{1}{2}}^k \frac{T_{i+1}^k-T_{i}^k}{\Delta z} \left. \right\rbrace.
\end{split}
\end{equation}
Analogously, according to \cref{fig:nTTMgrid} and \cref{eq:Ju-m}, the carrier current is:
\begin{equation}\label{eq:Jik}
\begin{split}
J_i^k= -D_i^k\Bigg[\frac{n_{i+1}^k-n_i^k}{\Delta z} + \frac{n_{i+\frac{1}{2}}^k}{k_B T_{i+\frac{1}{2}}^k}
\left\lbrace\left[H_{-\frac{1}{2}}^\frac{1}{2}(\eta_e)+H_{-\frac{1}{2}}^\frac{1}{2}(\eta_h)\right]_i^k \right\rbrace^{-1}
\frac{(E_g)_{i+1}^k-(E_g)_i^k}{\Delta z} \Bigg. +\\
\frac{n_{i+\frac{1}{2}}^k}{T_{i+\frac{1}{2}}^k}\left\lbrace
2\frac{\left[H_0^1(\eta_e)+H_0^1(\eta_h)\right]_{i+\frac{1}{2}}^k}
{\left[H_{\frac{1}{2}}^{-\frac{1}{2}}(\eta_e)+H_\frac{1}{2}^{-\frac{1}{2}}(\eta_h)\right]_{i+\frac{1}{2}}^k}
-\frac{3}{2} \right\rbrace \frac{T_{i+1}^k-T_i^k}{\Delta z} \Bigg. \Bigg]
\end{split}
\end{equation}
with
\begin{equation}\label{eq:Dik}
\begin{split}
D_i^k=\frac{k_B T_{i+\frac{1}{2}}^k}{q_e}
\frac{\mu_e \mu_h (H_\frac{1}{2}^0 \left( \eta_e \right))_{i+\frac{1}{2}}^k (H_\frac{1}{2}^0 \left( \eta_h \right))_{i+\frac{1}{2}}^k}
{\mu_e (H_\frac{1}{2}^0\left( \eta_e \right))_{i+\frac{1}{2}}^k+\mu_h ()H_\frac{1}{2}^0\left( \eta_h \right))_{i+\frac{1}{2}}^k}
\left[ H_{-\frac{1}{2}}^\frac{1}{2}\left( \eta_e \right) +  H_{-\frac{1}{2}}^\frac{1}{2}\left( \eta_h \right)\right]_{i+\frac{1}{2}}^k.
\end{split}
\end{equation}

Any function in between cells can be found by averaging:
\begin{equation}\label{eq:24-m}
A_{i+\frac{1}{2}}=\frac{1}{2}(A_i+A_{i+1}).
\end{equation}
The Fermi-Dirac integrals were calculated using GNU Scientific Library \cite{gsl} and stored in the tables in order
to speed up the calculations. $\frac{\partial\eta_c}{\partial n}$ can be found by taking the derivative of \cref{eq:eta_c-m} by the carrier density:
\begin{equation}\label{eq:25-m}
\left(\frac{\partial\eta_c}{\partial n}\right)=\frac{1}{2}\left(\frac{2\pi\hbar^2}{m_c^*k_BT_e}\right)^\frac{3}{2}\frac{1}{F_\frac{1}{2}(\eta_c)};
\end{equation}
$\frac{\partial\eta_c}{\partial T_e}$ can be found by taking the derivative of equation \cref{eq:eta_c-m} by the electronic temperature:
\begin{equation}\label{eq:26-m}
\left(\frac{\partial n_c}{\partial T_e}\right)=
-\frac{3}{\sqrt{2}}n\times(T_e)^{-\frac{5}{2}}
\left(\frac{\pi\hbar^2}{m_c^*k_B}\right)^\frac{3}{2} 
\frac{1}{F_\frac{1}{2}(\eta_c)}.
\end{equation}

The boundary conditions can be rewritten in the finite-difference form as follows:
\begin{equation}\label{eq:27-m}
\frac{n_1^{k+1}-n_1^k}{\Delta t}+
\frac{2J_1^k}{\Delta z}=
(S_n)_1^k-\gamma(n_1^k)^3+\delta_1^kn_1^k;
\end{equation}
\begin{equation}\label{eq:20-m-left}
(C_a)_1^k\frac{(T_a)_1^{k+1}-(T_a)_1^k}{\Delta t}=\frac{2}{(\Delta z)^2}(k_a)^k_{1+\frac{1}{2}}(T_{2}^k-T_{1}^k)+\frac{(C_{e-h})_1^k}{(\tau_{e-p})_1^k}\left[T_1^k-(T_a)_1^k\right] ;
\end{equation}
\begin{equation}\label{eq:28-m}
\frac{T_1^{k+1}-T_1^k}{\Delta t}=
(1-\psi)f_1^k+\psi f_1^{k+1}
\end{equation}
with
\begin{equation}\label{eq:29-m}
\begin{split}
(C_{e-h})_1^k f_1^k=&(S_u)_1^k-\frac{2W_1^k}{\Delta z}- 
\frac{(C_{e-h})_1^k}{(\tau_{e-p})_1^k}\left[T_1^k-(T_a)_1^k\right]- \\
&\left(\frac{\partial n}{\partial t}\right)_1^k\left\lbrace (E_g)_1^k+\frac{3}{2}k_BT_1^k 
\left[H_\frac{1}{2}^\frac{3}{2}(\eta_e)+H_\frac{1}{2}^\frac{3}{2}(\eta_h)\right]_1^k \right\rbrace- \\
&n_1^k\left\lbrace \left(\frac{\partial E_g}{\partial n}\right)_1^k \left(\frac{\partial n}{\partial t} \right)_1^k+\left(\frac{\partial E_g}{\partial T_a}\right)_1^k\left(\frac{\partial T_a}{\partial t}\right)_1^k   \right\rbrace- 
\\
&\frac{3}{2}k_BT_1^kn_1^k\left(\frac{\partial n}{\partial t} \right)_1^k \times
\left\lbrace \left(\frac{\partial\eta_e}{\partial n}
\right)_1^k \left[1-\left(H_{\frac{1}{2}}^{\frac{3}{2}}
(\eta_e)\right)_1^k \left(H_{\frac{1}{2}}^{-\frac{1}{2}}
(\eta_e)\right)_1^k\right]+ \right. \\
&\left. \left(\frac{\partial\eta_h}{\partial n} \right)_1^k
\left[1-\left(H_\frac{1}{2}^\frac{3}{2}(\eta_h)\right)_1^k
\left(H_\frac{1}{2}^{-\frac{1}{2}}(\eta_h)\right)_1^k
\right] 
\right\rbrace ;
\end{split}
\end{equation}
and on the other edge:
\begin{equation}\label{eq:30-m}
\frac{n_N^{k+1}-n_N^k}{\Delta t}-
\frac{2J_N^k}{\Delta z}=
(S_n)_N^k-\gamma(n_N^k)^3+\delta_N^kn_N^k;
\end{equation}
\begin{equation}\label{eq:20-m-left}
(C_a)_N^k\frac{(T_a)_N^{k+1}-(T_a)_N^k}{\Delta t}=\frac{-2}{(\Delta z)^2}(k_a)^k_{N-\frac{1}{2}}(T_{N}^k-T_{N-1}^k)+\frac{(C_{e-h})_N^k}{(\tau_{e-p})_N^k}\left[T_N^k-(T_a)_N^k\right] ;
\end{equation}
\begin{equation}\label{eq:31-m}
\frac{T_N^{k+1}-T_N^k}{\Delta t}=
(1-\psi)f_N^k+\psi f_N^{k+1}
\end{equation}
with
\begin{equation}\label{eq:32-m}
\begin{split}
(C_{e-h})_N^k f_N^k=&(S_u)_N^k+\frac{2W_N^k}{\Delta z}-
\frac{(C_{e-h})_N^k}{(\tau_{e-p})_N^k}\left[T_N^k-(T_a)_N^k\right]- \\
&\left(\frac{\partial n}{\partial t}\right)_N^k\left\lbrace (E_g)_N^k+\frac{3}{2}k_BT_N^k 
\left[H_\frac{1}{2}^\frac{3}{2}(\eta_e)+H_\frac{1}{2}^\frac{3}{2}(\eta_h)\right]_N^k \right\rbrace- \\
&n_N^k\left\lbrace \left(\frac{\partial E_g}{\partial n}\right)_N^k \left(\frac{\partial n}{\partial t} \right)_N^k+\left(\frac{\partial E_g}{\partial T_a}\right)_N^k\left(\frac{\partial T_a}{\partial t}\right)_N^k   \right\rbrace- \\
&\frac{3}{2}k_BT_N^kn_N^k\left(\frac{\partial n}{\partial t} \right)_N^k \times
\left\lbrace \left(\frac{\partial\eta_e}{\partial n}
\right)_N^k \left[1-\left(H_{\frac{1}{2}}^{\frac{3}{2}}
(\eta_e)\right)_N^k \left(H_{\frac{1}{2}}^{-\frac{1}{2}}
(\eta_e)\right)_N^k\right]+ \right. \\
&\left. \left(\frac{\partial\eta_h}{\partial n} \right)_N^k
\left[1-\left(H_\frac{1}{2}^\frac{3}{2}(\eta_h)\right)_N^k
\left(H_\frac{1}{2}^{-\frac{1}{2}}(\eta_h)\right)_N^k
\right] 
\right\rbrace .
\end{split}
\end{equation}
All the other equations and connections between variables at the boundaries stay the same and can be
straightforwardly obtained by substituting $i=1$ and $i=N$ into \cref{eq:19-m,eq:21-m,eq:22-m,eq:23-m,eq:25-m,eq:26-m}.

At the current time step $k$ we do not have any information about the following parameters from the future time step $k+1$:
\begin{equation} \label{eq:new-ts-list}
\begin{split}
&T_{i-1}^{k+1},T_i^{k+1},T_{i+1}^{k+1}, \left(\frac{\partial n}{\partial t}\right)_i^{k+1},
\left(\frac{\partial T_a}{\partial t}\right)_i^{k+1},
(S_u)_i^{k+1}, 
(C_{e-h})_i^{k+1}, J_i^{k+1}, \\
&\left(H_\frac{1}{2}^\frac{3}{2}(\eta_e)\right)_i^{k+1},
\left(H_\frac{1}{2}^\frac{3}{2}(\eta_h)\right)_i^{k+1},
\left(H_\frac{1}{2}^{-\frac{1}{2}}(\eta_e)\right)_i^{k+1},
\left(H_\frac{1}{2}^{-\frac{1}{2}}(\eta_h)\right)_i^{k+1}, \\
&\left(\frac{\partial \eta_e}{\partial n}\right)_i^{k+1},
\left(\frac{\partial \eta_h}{\partial n}\right)_i^{k+1},
\left(\frac{\partial \eta_e}{\partial T_e}\right)_i^{k+1},
\left(\frac{\partial \eta_h}{\partial T_e}\right)_i^{k+1}.
\end{split}
\end{equation}
Initially we set them all except the first three to be equal to the corresponding old values (at time step $k$):
\begin{equation}\label{eq:33-m}
\left(A_i^{k+1}\right)^{(0)}=A_i^k.
\end{equation}
Here $(0)$ means the $0^{th}$ step of the corrector. With this assumption, \cref{eq:21-m} becomes \begin{equation}\label{eq:35-m}
a_iT_{i-1}^{k+1}+b_iT_i^{k+1}+c_iT_{i+1}^{k+1}=r_i,
\end{equation}
which can be represented as a tridiagonal system of equations,
\begin{equation}\label{eq:34-m}
\begin{bmatrix}
b_1 & c_1 & ... &  & 0 \\[0.3em]
a_2 & b_2 & c_2 &  &\\[0.3em]
& a_3 & b_3 &  & ...\\[0.3em]
... &     & &...  & c_{N-1}\\[0.3em]
0   & 0   & ... & a_N & b_N \\
\end{bmatrix} 
\times
\begin{bmatrix}
T_1^{k+1}\\[0.3em]
T_2^{k+1} \\[0.3em]
...\\[0.3em]
T_{N-1}^{k+1}\\[0.3em]
T_N^{k+1}\\
\end{bmatrix} 
=
\begin{bmatrix}
r_1\\[0.3em]
r_2\\[0.3em]
...\\[0.3em]
r_{N-1}\\[0.3em]
r_N\\
\end{bmatrix} ,
\end{equation}
where
\begin{equation}\label{eq:36-m}
a_i=-\psi \frac{\Delta t}{\Delta z(C_{e-h})_i^{k+1}}
\left(k_B[H_0^1(\eta_e)+H_0^1(\eta_h)]_{i-\frac{1}{2}}^{k+1}\times J_{i-1}^{k+1}+(k_e+k_h)_{i-\frac{1}{2}}^{k+1}
/\Delta z\right),
\end{equation}
\begin{equation}\label{eq:37-m}
\begin{split}
b_i=1-&\psi \frac{\Delta t}{(C_{e-h})_i^{k+1}}\times  
\Bigg(-\frac{k_B}{\Delta z}
\left[H_0^1(\eta_e)+H_0^1(\eta_h)\right]_{i+\frac{1}{2}}^{k+1}\times J_i^{k+1}+ \\
&\frac{k_B}{\Delta z}
\left[H_0^1(\eta_e)+H_0^1(\eta_h)\right]_{i-\frac{1}{2}}^{k+1}\times J_{i-1}^{k+1}- \Bigg. \\
&\left[(k_e+k_h)_{i-\frac{1}{2}}^{k+1}+
(k_e+k_h)_{i+\frac{1}{2}}^{k+1}\right]/
(\Delta z)^2-(C_{e-h})_i^{k+1}/(\tau_{e-p})_i^{k+1}- \\
&\frac{3}{2}(\frac{\partial n}{\partial t})_i^{k+1}
k_B\left(H_\frac{1}{2}^\frac{3}{2}(\eta_e)+
H_\frac{1}{2}^\frac{3}{2}(\eta_h)\right)_i^{k+1}
-\frac{3}{2}k_Bn_i^{k+1}\left(\frac{\partial n}{\partial t} \right)_i^{k+1} \times\\
&\left\lbrace \left(\frac{\partial\eta_e}{\partial n}
\right)_i^{k+1} \left[1-\left(H_{\frac{1}{2}}^{\frac{3}{2}}
(\eta_e)\right)_i^{k+1} \left(H_{\frac{1}{2}}^{-\frac{1}{2}}
(\eta_e)\right)_i^{k+1}\right]+ \right. \\
&\left. \left(\frac{\partial\eta_h}{\partial n} \right)_i^{k+1}
\left[1-\left(H_\frac{1}{2}^\frac{3}{2}(\eta_h)\right)_i^{k+1}
\left(H_\frac{1}{2}^{-\frac{1}{2}}(\eta_h)\right)_i^{k+1}
\right] 
\right\rbrace  \Bigg. \Bigg) ,
\end{split}
\end{equation}
\begin{equation}\label{eq:38-m}
c_i=\psi \frac{\Delta t}{\Delta z(C_{e-h})_i^{k+1}}
\left(k_B[H_0^1(\eta_e)+H_0^1(\eta_h)]_{i+\frac{1}{2}}^{k+1}\times J_i^{k+1}-(k_e+k_h)_{i+\frac{1}{2}}^{k+1}
/\Delta z\right),
\end{equation}
\begin{equation}\label{eq:39-m}
\begin{split}
r_i=T_i^k+&(1-\psi)\frac{\Delta t}{(C_{e-h})_i^k}f_i^k+
\psi\frac{\Delta t}{(C_{e-h})_i^{k+1}} \times\\
&\Bigg(S_i^{k+1}-(E_g)_{i+\frac{1}{2}}^{k+1}\times
\frac{J_i^{k+1}}{\Delta z}+
(E_g)_{i-\frac{1}{2}}^{k+1}\times 
\frac{J_{i-1}^{k+1}}{\Delta z}+ \Bigg. \\ 
&\frac{(C_{e-h})_i^{k+1}}{(\tau_{e-p})_i^{k+1}}
(T_a)_i^{k+1}-(\frac{\partial n}{\partial t})_i^{k+1}
(E_g)_i^{k+1}- \\ 
&n_i^{k+1}\left\lbrace \left(\frac{\partial E_g}{\partial n}\right)_i^{k+1} \left(\frac{\partial n}{\partial t}\right)_i^{k+1}+\left(\frac{\partial E_g}{\partial T_a}\right)_i^{k+1} \left(\frac{\partial T_a}{\partial t}\right)_i^{k+1}  \right\rbrace \Bigg. \Bigg)
\end{split}
\end{equation}
for $i=2,...,N-1$, and the boundary conditions are:
\begin{equation}\label{eq:40-m}
\begin{split}
b_1=1-&\psi \frac{\Delta t}{(C_{e-h})_1^{k+1}}\times 
\Bigg(-\frac{k_B}{\Delta z}
\left[H_0^1(\eta_e)+H_0^1(\eta_h)\right]_1^{k+1}\times J_1^{k+1}- \\
&\frac{(k_e+k_h)_\frac{3}{2}^{k+1}}{(\Delta z)^2}-(C_{e-h})_1^{k+1}/(\tau_{e-p})_1^{k+1}- \\
&\frac{3}{2}k_B(\frac{\partial n}{\partial t})_1^{k+1}
\left(H_\frac{1}{2}^\frac{3}{2}(\eta_e)+
H_\frac{1}{2}^\frac{3}{2}(\eta_h)\right)_1^{k+1}
-\frac{3}{2}k_Bn_1^{k+1}\left(\frac{\partial n}{\partial t} \right)_1^{k+1} \times\\
&\left\lbrace \left(\frac{\partial\eta_e}{\partial n}
\right)_1^{k+1} \left[1-\left(H_{\frac{1}{2}}^{\frac{3}{2}}
(\eta_e)\right)_1^{k+1} \left(H_{\frac{1}{2}}^{-\frac{1}{2}}
(\eta_e)\right)_1^{k+1}\right]+ \right. \\
&\left. \left(\frac{\partial\eta_h}{\partial n} \right)_1^{k+1}
\left[1-\left(H_\frac{1}{2}^\frac{3}{2}(\eta_h)\right)_1^{k+1}
\left(H_\frac{1}{2}^{-\frac{1}{2}}(\eta_h)\right)_1^{k+1}
\right] 
\right\rbrace  \Bigg. \Bigg) ,
\end{split}
\end{equation}
\begin{equation}\label{eq:41-m}
c_1=-\psi \frac{2\Delta t}{\Delta z(C_{e-h})_1^{k+1}}
\left(-k_B[H_0^1(\eta_e)+H_0^1(\eta_h)]_{\frac{3}{2}}^{k+1}\times J_1^{k+1}+(k_e+k_h)_\frac{3}{2}^{k+1}
/\Delta z\right) ,
\end{equation}
\begin{equation}\label{eq:42-m}
\begin{split}
r_1=&T_1^k+(1-\psi)\Delta tf_1^k+
\psi\frac{\Delta t}{(C_{e-h})_1^{k+1}} \times\\
&\Bigg(S_1^{k+1}-2(E_g)_\frac{3}{2}^{k+1}\times
\frac{J_1^{k+1}}{\Delta z}+
\Bigg.  
\frac{(C_{e-h})_1^{k+1}}{(\tau_{e-p})_i1^{k+1}}
(T_a)_1^{k+1}-\left(\frac{\partial n}{\partial t}\right)_1^{k+1}
(E_g)_1^{k+1}- \\ 
&n_1^{k+1}\left\lbrace \left(\frac{\partial E_g}{\partial n}\right)_1^{k+1} \left(\frac{\partial n}{\partial t}\right)_1^{k+1}+\left(\frac{\partial E_g}{\partial T_a}\right)_1^{k+1} \left(\frac{\partial T_a}{\partial t}\right)_1^{k+1}  \right\rbrace \Bigg. \Bigg)
\end{split}
\end{equation}
and 
\begin{equation}\label{eq:43-m}
\begin{split}
a_N=&-\psi \frac{\Delta t}{\Delta z(C_{e-h})_N^{k+1}} \times\\
&\Big(2k_B[H_0^1(\eta_e)+H_0^1(\eta_h)]_{N-\frac{1}{2}}^{k+1}\times J_{N-1}^{k+1}+ \Big. 
(k_e+k_h)_{N-\frac{1}{2}}^{k+1}/\Delta z \Big. \Big),
\end{split}
\end{equation}
\begin{equation}\label{eq:44-m}
\begin{split}
b_N=1-&\psi \frac{\Delta t}{(C_{e-h})_N^{k+1}}\times 
\Bigg(-2\frac{k_B}{\Delta z}
\left[H_0^1(\eta_e)+H_0^1(\eta_h)\right]_{N-\frac{1}{2}}^{k+1}\times J_{N-1}^{k+1}- \\
&\frac{2(k_e+k_h)_{N-\frac{1}{2}}^{k+1}}{(\Delta z)^2}-(C_{e-h})_N^{k+1}/(\tau_{e-p})_N^{k+1}- \\
&\frac{3}{2}k_B\left(\frac{\partial n}{\partial t}\right)_N^{k+1}
\left(H_\frac{1}{2}^\frac{3}{2}(\eta_e)+
H_\frac{1}{2}^\frac{3}{2}(\eta_h)\right)_N^{k+1}
-\frac{3}{2}k_Bn_N^{k+1}\left(\frac{\partial n}{\partial t} \right)_N^{k+1} \times\\
&\left\lbrace \left(\frac{\partial\eta_e}{\partial n}
\right)_N^{k+1} \left[1-\left(H_{\frac{1}{2}}^{\frac{3}{2}}
(\eta_e)\right)_N^{k+1} \left(H_{\frac{1}{2}}^{-\frac{1}{2}}
(\eta_e)\right)_N^{k+1}\right]+ \right. \\
&\left. \left(\frac{\partial\eta_h}{\partial n} \right)_N^{k+1}
\left[1-\left(H_\frac{1}{2}^\frac{3}{2}(\eta_h)\right)_N^{k+1}
\left(H_\frac{1}{2}^{-\frac{1}{2}}(\eta_h)\right)_N^{k+1}
\right] 
\right\rbrace  \Bigg. \Bigg) ,
\end{split}
\end{equation}
\begin{equation}\label{eq:45-m}
\begin{split}
r_N=&T_N^k+(1-\psi)\Delta t f_N^k+
\psi\frac{\Delta t}{(C_{e-h})_N^{k+1}} \times\\
&\Bigg(S_N^{k+1}+2(E_g)_{N-\frac{1}{2}}^{k+1}\times
\frac{J_{N-1}^{k+1}}{\Delta z}+
\Bigg. 
\frac{(C_{e-h})_N^{k+1}}{(\tau_{e-p})_N^{k+1}}
(T_a)_N^{k+1}-\left(\frac{\partial n}{\partial t}\right)_N^{k+1}
(E_g)_N^{k+1}- \\ 
&n_N^{k+1}\left\lbrace \left(\frac{\partial E_g}{\partial n}\right)_N^{k+1} \left(\frac{\partial n}{\partial t}\right)_N^{k+1}+\left(\frac{\partial E_g}{\partial T_a}\right)_N^{k+1} \left(\frac{\partial T_a}{\partial t}\right)_N^{k+1}  \right\rbrace \Bigg. \Bigg) .
\end{split}
\end{equation}
Such a system can be resolved with respect to $\left\lbrace T_i^{k+1} \right\rbrace_{i=1}^N$ by using the well-known tridiagonal matrix algorithm \cite{thomas1949elliptic}.

We denote the electronic temperature calculated with assumption (\ref{eq:33-m}) as $(T_i^{k+1})^{(1)}$, showing with ‘$(1)$’
the first correction step. This result allows to calculate the corrected new values of $T_a$, $n$ (from \cref{eq:19-m} and \cref{eq:20-m}
respectively), and those in the list (\ref{eq:new-ts-list}):
\begin{equation}\label{eq:46-m}
(A_i^{k+1})^{(1)}=A_i(T=[T_i^{k+1}]^{(1)}).
\end{equation}
In turn, these values allow to calculate $(T_i^{k+1})^{(2)}$ from \cref{eq:34-m} and so on. Owing to its similarity with predictor-corrector methods, we call it “predictor-corrector” algorithm. With this approach, \cref{eq:21-m} can be rewritten in the following form:
\begin{equation}\label{eq:47-m}
\begin{split}
(a_i)^{(l)}(T_{i-1}^{k+1})^{(l+1)}+(b_i)^{(l)}(T_i^{k+1})^{(l+1)}+(c_i)^{(l)}(T_{i+1}^{k+1})^{(l+1)}=(r_i)^{(l)},\\ l=0, 1,\dots,
\end{split}
\end{equation}
where index $(l)$ shows the current step of correction and $(l)=(0)$ means the value is old, i.e., taken at time step $k$. This procedure continues until the difference between two last corrected values of electronic temperature is less than the demanded precision:
\begin{equation}\label{eq:48-m}
\sum_{i=1}^N\left[(T_i^{k+1})^{(l+1)}-(T_i^{k+1})^{(l)}\right]<\varepsilon .
\end{equation}
For the chosen precision of $\varepsilon$ = \SI{10^{-6}}{K}, it takes around \num{300} corrections to reach it during the laser pulse action, whereas when the laser is ended, \num{5} corrections is usually enough.

\section{Calculation example}
\label{sec:sol-alg-results}
\begin{figure}
\centering
	\includegraphics[width=\textwidth]{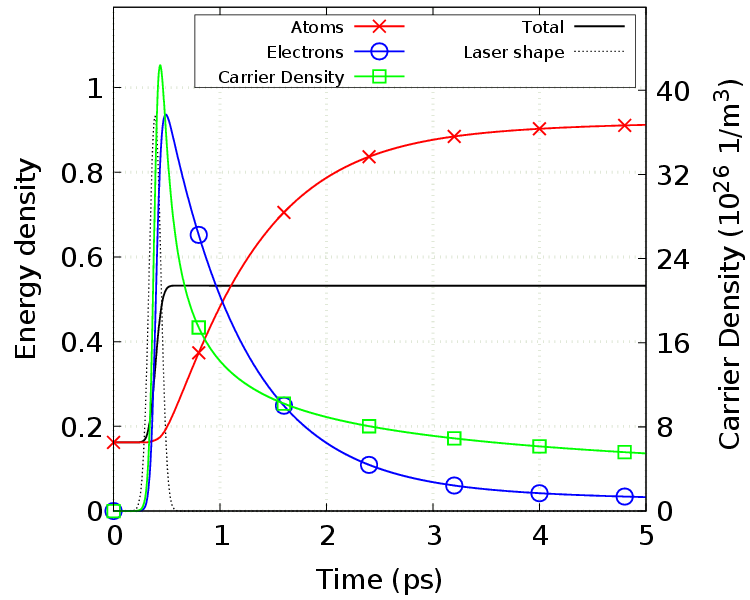}
	
	\caption{Electron/lattice energy densities (divided by the melting energy density) and carrier density dynamics, according to the \textit{n}TTM model, at the surface of silicon target of \SI{800}{nm} width, followed by the \SI{130}{\femto\second} laser pulse at the incident fluence of \SI{0.26}{J/cm^2}. The total energy density, averaged through the whole sample, is shown with black solid line. The laser pulse shape, shown with black dotted line, is not in scale.} 
	\label{fig:nTTMres-mesh}
\end{figure}

As an example of application of our algorithm to the described system of equations \cref{eq:dn_dt-m,eq:dTe_dt-m,eq:dTa_dt-m},
we perform the simulations of 800-nm-thick silicon target's response to ultrashort laser pulse irradiation. The parameters of the irradiation are \SI{130}{\femto\second} duration, \SI{800}{\nano\meter} wavelength, and \SI{0.26}{J/cm^2} incident fluence. For these conditions, the experimental melting threshold fluence is \SI{0.27}{J/cm^2} \cite{bonse2004modifying}, which is in agreement with the result of the \textit{n}TTM model \cite{lipp2015thesis}. The value of fluence is chosen to be just below the melting threshold, providing the applicability of this simple model in the absence of phase transition processes. The sample was divided into \SI{160}{cells} according to \cref{fig:nTTMgrid}. In \cref{fig:nTTMres-mesh} we show the dynamics of electron-hole carrier density and electronic and atomic energy densities at the silicon surface. The shown energy density is scaled to the melting energy density, which is found to be \SI{3.86\times10^9}{J/m^3}, according to the simulations. Though \cref{eq:dTe_dt-m,eq:dTa_dt-m} are written in terms of temperature of carriers and atoms, we plotted the corresponding energy densities instead, because their dynamics represents the energy flow between the subsystems and allows plotting the same scale for electrons and atoms, whereas electronic temperature is much higher than the atomic one (see also Fig. 2 in \cite{lipp2014atomistic}). In addition, this choice provides a possibility to show the energy conservation with the total average energy density of the sample (shown with black solid line).

The initial increase in the carrier density followed by the laser pulse is connected to the excitation of new carriers by one- and two-photon absorption processes. With time, the increase changes to the decay due to strong Auger recombination and diffusion processes. The strong peak in the electronic energy density is mostly connected to the free-carrier absorption. Finally, the thermal energy from electron-hole carriers is transferred to the atomic subsystem of the sample leading to gradual increase in the lattice energy density upon the electron-phonon equilibration.

\section{Precision, stability and calculation speed}
\label{sec:sol-alg-discuss}
In case of the explicit scheme, a good guess for the time step requirement can be obtained from the von Neumann stability criterion \cite{isaacson2012analysis}, $\Delta t\le \frac{(\Delta x)^2}{2D_{th}}$, where $D_{th}$ is thermal  diffusion coefficient, which is proportional to thermal conductivity $k_e$ and inversely proportional to the carrier heat capacity $C_{e-h}$. Under initial (prepulse) conditions, in the absence of free carriers, the latter tends to vanish (see \cref{eq:C_eh-m}), whereas the former is limited (see \cref{Tab:ModelParam-m}). After the laser irradiation starts, quick increase of $T_e$ at initially low $n$ (see also Fig. 2 in \cite{lipp2014atomistic}) leads to an abrupt increase of $D_{th}$, which influences the von Neumann stability criterion and limits the maximum possible time step for explicit integration methods. Consequently, if one applies an explicit finite-difference scheme for the numerical solution, the stability of \cref{eq:dTe_dt-m} limits the maximum possible time step to \SI{1\times10^{-19}}{\second} for the discretization of \SI{40}{cells} \cite{chen2005numerical}.

\begin{figure}
\centering
	\includegraphics[width=\textwidth]{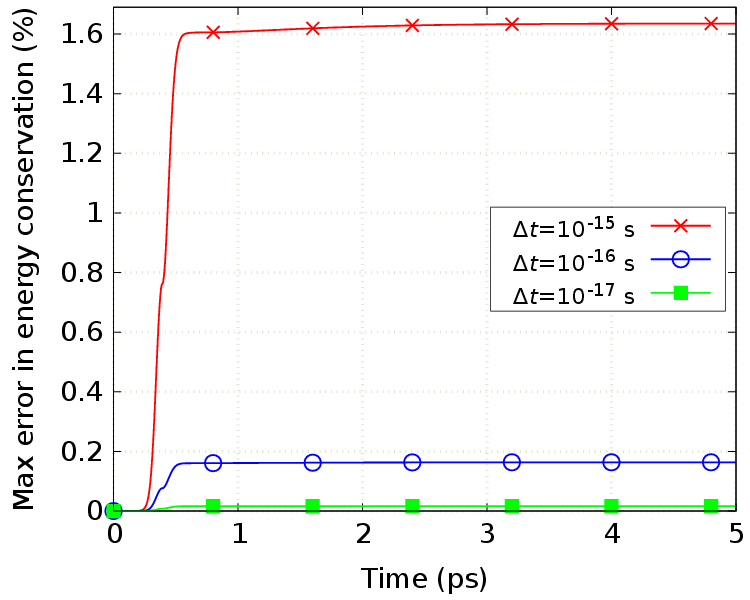}
	
	\caption{Maximum relative error in energy conservation as a function of simulation time for three different time steps. The quantity is calculated from the difference between total absorbed fluence and the energy in the system.} 
	\label{fig:implic-err}
\end{figure}

In contrast, the proposed semi-implicit numerical integration scheme provides a stable solution for time step as high as \SI{1\times10^{-16}}{\second} with the energy
conservation about \SI{0.16}{\%} per simulation, \cref{fig:implic-err}. At the same time, in case the calculation speed is critical, increasing the time step even higher is possible: \SI{1\times10^{-15}}{\second} provides the energy conservation within \SI{1.6}{\%}. Thus, the increase in the calculation speed of up to $10^5$ times has been achieved, compared with the explicit finite-difference integration scheme \cite{chen2005numerical}. Unfortunately, a mathematical error in ref. \cite{chen2005numerical} (specifically, in equations (18) and (19)) did not allow us to directly compare the results.

The time step is of course limited by all the characteristic times of the involved physical processes, such as laser pulse duration, electron-phonon coupling time, and carrier recombination time. As long as it is much smaller than those mentioned above, the presented integration scheme is tested to be unconditionally stable.

This approach has been successfully applied earlier in order to investigate and improve the presented \textit{n}TTM model \cite{ramer2014laser}. The atomistic-continuum model, describing the dynamics of gold targets under the ultrashort-pulse lasers, also benefited from using the presented approach \cite{ivanov2014molecular}. In our work \cite{lipp2014atomistic}, we used the described scheme for the solution of the continuum part of the atomistic-continuum model MD-\textit{n}TTM. The high speed and precision of the scheme allowed to significantly decrease the computational costs of the corresponding simulations.

In the mentioned applications, the corresponding system was solved in 1D, based on the assumption of wide laser spot in comparison with the lateral sizes of the computational setup. Whenever it is not the case, one needs to solve the corresponding problem (the vector system of \cref{eq:dn_dt-m,eq:dTe_dt-m,eq:dTa_dt-m}) in 2D or 3D case. According to the ref. \cite{wang20033d}, the diffusion equation in 3D case can be solved in 3 subsequent steps, each of which involves implicit solution in only one direction ($X$, $Y$, or $Z$) and explicit scheme in the other two directions. In other words, one can use 1D implicit scheme three times: for $X$, $Y$, and $Z$ directions separately and consequently. This approach is called alternating direction implicit method (ADI) and is also widely applied for the corresponding 2D problems \cite{peaceman1955numerical}.
Therefore, with appropriate modifications, the presented scheme should be applicable for the considered problem in 2D and 3D cases as well.

\section{Conclusion}
\label{sec:conclude}
We proposed the semi-implicit integration scheme for the solution of diffusion-like nonlinear equations. The scheme is based on the Crank-Nicolson finite-difference integration method, modified with a predictor-corrector algorithm, according to \cref{eq:47-m}. The modification resulted in a possibility to solve nonlinear diffusion equations with high stability and precision.

In the presented example of the scheme application, we reached the speed up of the calculations (by the
increase of the integration time step) by $10^4$ times compared with the explicit scheme, keeping the error in
energy conservation below \SI{0.2}{\%}. This error increases linearly with the time step. The algorithm is applicable in case the time step is much smaller than all the characteristic times of the involved physical processes. The existing applications that use the proposed scheme are mentioned and the possible extension for 2D and 3D cases is suggested.

\section*{Acknowledgements}

This work was supported by the DFG grants IV 122/1-1, IV 122/2, and RE 1141/15-1. The authors acknowledge Markus Nie{\ss}en for the assistance in the development of the described solution scheme. The work was partly conducted at the Institute for Laser Technology (ILT), RWTH, Aachen, Germany, department of "Nonlinear Dynamics of Laser Processing (NLD)", in the group of Prof. Dr. Wolfgang Schulz.

\bibliography{Lipp_numerical_scheme.bib}

\end{document}